\documentstyle[emulapj,epsfig]{article}

\font\tenbg=cmmib10 at 10pt

\def \rvecphi{{\hbox{\tenbg\char'036}}}

\begin{document}
\title{Poynting Jets from Accretion Disks}

\author{
R.V.E. Lovelace \altaffilmark{1}, H. Li \altaffilmark{2}, A.V.
Koldoba \altaffilmark{3}, G.V. Ustyugova \altaffilmark{4},
M.M. Romanova \altaffilmark{1}}

\altaffiltext{1}{Department of Astronomy,
Cornell University, Ithaca, NY 14853-6801;
RVL1@cornell.edu; Romanova@astrosun.tn.cornell.edu}

\altaffiltext{2}{Theoretical Astrophysics, T-6, MS B288,
Los Alamos National Laboratory, Los Alamos, NM 87545; 
HLi@lanl.gov}

\altaffiltext{3}{Institute of
Mathematical Modelling,
Russian Academy of Sciences, Moscow,
Russia, 125047; Koldoba@spp.keldysh.ru}

\altaffiltext{4}{Keldysh Institute of
Applied Mathematics, Russian Academy
of Sciences, Moscow, Russia, 125047,
Ustyugg@spp.Keldysh.ru}

\begin{abstract}

    We give further considerations
on the problem of the evolution
of a coronal, force-free magnetic field which
threads a differentially rotating, conducting
Keplerian disk, extending the work of
Li {\it et al.} (2001).
    This situation is described 
by the force-free 
Grad-Shafranov (GS) equation
for the flux function $\Psi(r,z)$
which labels the poloidal field lines
(in  cylindrical coordinates).
   The GS equation
involves a function
$H(\Psi)$ describing the
distribution of poloidal current 
which is determined by the differential
rotation or {\it twist} of 
the disk which increases linearly with
time.
  We numerically solve the GS equation
in a sequence of volumes of increasing
size corresponding to the
expansion of the outer perfectly
conducting boundaries at ($R_{m},~Z_{m}$).
  The outer boundaries model the
influence of an external non-magnetized
plasma.
  The sequence of GS solutions provides
a model for the dynamical evolution of the
magnetic field in response to (1) the 
increasing twist
of the disk and (2) the pressure of external
plasma.
    We find solutions with {\it magnetically
collimated} Poynting jets 
where there is a {\it continuous} outflow
of energy, angular momentum, and toroidal magnetic
flux from the disk into the external space.  
   This behavior contradicts
the commonly accepted ``theorem'' 
of Solar plasma physics
that the motion
of the footpoints of a magnetic
loop structure leads
to a stationary magnetic
field configuration 
with zero power, angular momentum, 
and flux outflows.

   In addition we discuss 
magnetohydrodynamic (MHD) simulations 
which show quasi-stationary collimated
Poynting jets similar to our Grad-Shafranov
solutions. 
   In contrast with the Grad-Shafranov
solutions, the simulations show  a steady
uncollimated hydromagnetic 
(non force-free) outflow
from the outer part of the disk.
   The Poynting jets are of interest
for the understanding of
the jets from active galactic
nuclei, microquasars, and possibly 
gamma ray burst sources.

\end{abstract}

\section{Introduction}

Highly-collimated,
oppositely directed jets are
observed in
active galaxies and 
quasars (see for example 
Bridle \& Eilek 1984),
and in
old compact stars in binaries 
(Mirabel \& Rodriguez 1994;
Eikenberry {\it et al.} 1998).
   Further, well collimated
emission line jets are
seen in young stellar
objects (Mundt 1985;  B\"uhrke, Mundt,
\& Ray 1988).
Recent work favors
 models where twisting of an
ordered magnetic field
threading an accretion
disk acts to magnetically
accelerate the jets (Meier, Koide, \& Uchida 2001).
   There are two regimes:
 (1)  the {\it hydromagnetic regime}, 
where energy and angular 
momentum are carried by both
the electromagnetic field and
the kinetic flux of matter, which 
is relevant to the jets from 
young stellar objects;
and (2) the {\it Poynting flux regime}.
where  energy and angular
from the disk are carried predominantly by the
electromagnetic field, which is relevant
to extra-galactic and microquasar jets,
and possibly to gamma ray burst sources.

    Stationary 
Poynting flux dominated
jets have been found in
axisymmetric MHD simulations of
the opening of  magnetic loops 
threading a Keplerian disk (Romanova {\it et al.} (1998); 
Ustyugova {\it et al.} 2000).
    Theoretical studies 
have developed  models for Poynting outflows
from accretion disks (Lovelace, Wang, \& Sulkanen 1987;
Colgate \& Li 1998). 
    The present work represents a continuation
of the study by Li {\it et al.} (2001) where
self-consistent,
axisymmetric, non-relativistic solutions of
the  Grad-Shafranov equation are calculated
inside given conducting boundaries.
   The solutions are self-consistent
in the respect that the twist of each field
line is that due to the differential rotation
of a Keplerian disk.

\begin{figure*}
\caption{
 ``Initial'' dipole-like
vacuum 
magnetic field.
   In this and subsequent plots,
$r$ and $z$ are measured in units
of the radius $r_0$ of the $O-$point
in the disk plane.
   The solid lines are the magnetic
field lines for the case where
the flux function on the disk surface is
$\Psi(r,z=0)=
a^3r^2 K/(a^2+r^2)^{3/2}$ with
$a=r_0/\sqrt{2}$.
    In this and subsequent
plots $\Psi$ is measured in units
of $\Psi_0=r_0^2 K/3^{3/2}$.  
Note that $B_z(0) = 
6\sqrt{3}\Psi_0/r_0^2 
\approx 10.4\Psi_0/r_0^2$.
   The dashed lines are the field lines
for the case where the outer boundaries
($R_{m}=10,~Z_{m}=10$)
are perfectly conducting; 
for this case
$\Psi(r,z=0) \rightarrow 
\Psi(r,0)[1-(r/r_0-1)^2/81]$ so that the
$O-$point is still at $r_0$ and $\Psi_0$
is unchanged.
     Because of  axisymmetry
and reflection symmetry about the $z=0$ plane,
the field need be shown in only one 
quadrant.
}
\end{figure*}

  In \S 2 we summarize the theory, and
in \S 3 we discuss the different types
of solutions found.  In \S 4 we develop
an analytic model for Poynting jets.
In \S 5, we discuss the consequences of
expanding the boundaries.
   Sections 6 - 9 discuss the relevant
conservation laws.  
  In \S 10 we compare the Poynting
outflows with centrifugally launched
winds.  
     In \S 11  we discuss the collapse
of the inner part of the disk due  to
angular momentum outflow to Poynting
jets.
     In \S 12 we discuss conditions
for occurrence of Poynting jets including
the influence of the kink instability.
    In \S 13, we give results of MHD
simulations which give Poynting jets
similar to our Grad-Shafranov solutions.
    Section 14 gives conclusions from
this work.

\section{Theory of Poynting Outflows}

    Here, we consider further the
theory of Poynting outflows (Li {\it et al.} 2001).
   We assume  that magnetic field loops threads
a differentially rotating highly
conducting Keplerian
accretion disk at some initial time $t=0$.
   Above the disk we assume a ``coronal''
or force-free magnetic field in
the non-relativistic limit.
   This situation is described 
by the force-free 
Grad-Shafranov (GS) equation
for the flux function $\Psi(r,z)$
which labels the poloidal field lines
(in  cylindrical coordinates).
   The GS equation
involves a function
$H(\Psi)$ describing the
distribution of poloidal current 
which is determined by the differential
rotation or {\it twist} of 
the disk which increases linearly with
time.
  We numerically solve the GS equation
in a sequence of volumes of increasing
size corresponding to the
expansion of the outer perfectly
conducting boundaries at ($R_{m},~Z_{m}$).
  The outer boundaries model the
influence of an external non-magnetized
plasma.
  The sequence of GS solutions provides
a model for the dynamical evolution of the
magnetic field in response to (1) the 
increasing twist
of the disk and (2) the pressure of external
plasma.

   Cylindrical $(r,\phi,z)$
coordinates are used and 
axisymmetry is assumed
   Thus the magnetic field has
the form
$ {\bf B}~ = {\bf B}_p +
B_\phi \hat{\rvecphi~}~,
$
with
$
{\bf B}_p = B_r{\hat{\bf r}}+
B_z \hat{\bf z}~.
$
We can write
$
B_r =
-(1 / r)(\partial \Psi
/ \partial z),
B_z =(1 / r)
(\partial \Psi / \partial r),
$
where $\Psi(r,z) \equiv r A_\phi(r,z)$.
   In the 
force-free limit,
the magnetic energy density ${\bf B}^2/(8\pi)$
is much larger than the kinetic or thermal
energy densities; that is, the flow speeds are
sub-Alfv\'enic,
${\bf v}^2 \ll {v}_A^2 = {\bf B}^2/4\pi \rho$,
where $v_A$ is the Alfv\'en velocity.
In this limit,
$
{ 0} \approx {\bf J \times B}$;
   therefore, ${\bf J} = \lambda {\bf B}$
(Gold \& Hoyle 1960).
  Because ${\bf \nabla \cdot J}=0$,
$({\bf B \cdot \nabla}) \lambda =0$
and consequently $\lambda = \lambda(\Psi)$,
as  well-known.
   Thus, Amp\`ere's equation becomes
$
{\bf \nabla \times B} = {4 \pi \lambda(\Psi)} {\bf B}/c
$.
   The $r$ and $z$ components of Amp\`ere's
equation  imply
$$
 rB_\phi = H(\Psi)~,\quad 
{\rm and}\quad {dH(\Psi) / d\Psi} 
={4\pi}
\lambda(\Psi)/c,
$$
where $H(\Psi)$ is another function of
$\Psi$.   
   Thus, $H(\Psi)=$ const are lines of
constant poloidal current density;
${\bf J}_p=(c / 4\pi)
(dH / d\Psi){\bf B}_p$ so that
$({\bf J}_p \cdot{\bf \nabla}) H=0$. 
   The toroidal component of Amp\`ere's
equation gives
\begin{equation}
\Delta^\star \Psi = -
H(\Psi) {d H(\Psi) \over d\Psi}~,
\end{equation}
with $ \Delta^\star \equiv
{\partial^2 / \partial r^2}
-(1/r)(\partial / \partial r)
+{\partial^2 / \partial z^2},
$
which is the Grad-Shafranov equation for
$\Psi$ (see e.g. Lovelace {\it et al.} 1987;
Li {\it et al.} 2001).

   Amp\`ere's law gives
$\oint d{\bf l}\cdot {\bf B}
=(4\pi/c)\int d{\bf S}\cdot {\bf J}$,
so that
$rB_\phi(r,z)=H(\Psi)$ is $(2/c)$ times 
the current flowing through a circular
area of radius $r$ (with normal $\hat{\bf z}$) 
labeled by $\Psi(r,z)$= const. 
  Equivalently, $-H[\Psi(r,0)]$ 
is  $(2/c)\times$
the current flowing into the
area of the disk $\leq r$.
   For all cases studied here,
$-H(\Psi)$ has a maximum so that
the total current flowing into
the disk for $r\leq r_m$ is
$
 I_{tot} = (2 / c)(-H)_{max},
$
where $r_m$ is such that
$-H[\Psi(r_m,0)]=(-H)_{max}$
so that $r_m$
is less than 
the radius of the $O-$point, $r_0$.
  The same total current $I_{tot}$
flows out of the region of the disk
$r=r_m$ to $r_0$.

     We consider
an {\it initial value problem} where the
disk at $t=0$ is threaded by a 
dipole-like poloidal magnetic field.
     The form of $H(\Psi)$
is then determined
by the differential rotation of the
disk:
   The azimuthal {\it twist} of a given field
line going from an inner footpoint
at $r_1$ to an outer footpoint at $r_2$
is fixed by the differential rotation
of the disk.
  The field line slippage speed through
the disk due to the disk's finite
magnetic diffusivity is estimated
to be negligible compared with the
Keplerian velocity.
  For a given field line
we have $rd\phi/B_\phi = ds_p/B_p$,
where $ds_p \equiv \sqrt{dr^2+dz^2}$ is the
poloidal arc length along the field
line, and
$B_p \equiv \sqrt{B_r^2+B_z^2}$.
   The total twist of a field line
loop is
\begin{equation}
\label{j31}
\Delta \phi(\Psi) =
\int_1^2 ds_p ~{-B_\phi \over r B_p}
=-H(\Psi) \int_1^2 {ds_p \over r^2 B_p}~,
\end{equation}
with the sign included to give
$\Delta \phi >0$.
  For a Keplerian disk around
an object of mass $M$ the angular
rotation is $\omega_K = \sqrt{G M/r^3}$
so that
the field line twist after a time $t$ is
\begin{equation}
\label{j32}
\Delta \phi(\Psi)
=   \omega_0 ~t
\left[\left({r_0\over r_1}\right)^{3/2} -
\left({r_0\over r_2}\right)^{3/2}\right]
= (\omega_0~t) ~{\cal F}(\Psi/\Psi_0)
\end{equation}
where $r_0$ is the radius of the
$O-$point,
$\omega_0\equiv\sqrt{GM/r_0^3}$,
and ${\cal F}$ is a dimensionless
function (the quantity in the
square brackets).  
  The $O-$point is the point in
the midplane of the disk encircled by
the poloidal magnetic field lines;
at this point ${\bf B}_p=0$.
  At sufficiently small
$r_1$ one  reaches the inner
radius of the disk $r_i~(\ll r_0)$
where we assume $\omega_K$ saturates
at the value $\omega_{Ki}=\sqrt{GM/r_i^3}$.
    For the dipole-like field
of Figure 1, ${\cal F}\approx 3^{9/8}
(\Psi_0/\Psi)^{3/4}$
for $\Psi/\Psi_0 \ll 1$, while
${\cal F} 
\approx  3.64(1-\Psi/\Psi_0)^{1/2}$
for $1-\Psi/\Psi_0 \ll 1$.

  For an accretion
disk around a massive
black hole $M=M_810^8M_\odot$ in
the nucleus of galaxy, 
the twist is
$ T= (t/3.17{\rm d})(r_0/10^{15}{\rm cm})^{3/2}
/\sqrt{M_8}$.  
   The inner radius of
the disk is $r_i\approx M_8 9 \times 10^{13}$ cm
for a Schwarzschild black hole.

\begin{figure*} 
\caption{
Summary of dependences
of numerical solutions  of
equation (1) on the twist
$T=\omega_0t$.
  For these solutions
$H(\Psi)$ satisfies equations (2) and (3). 
   Here, $W_m$ is the magnetic
energy (with $W_{m0}\approx 0.57$ the energy for
zero twist),  $\Phi_t$ is the toroidal
flux, and $(-H)_{max}$ is the total
poloidal current (times $c/2$).
    The solutions are inside
a conducting box with $R_{m}=10=
Z_{m}$ for an initial dipole-like
field with $O-$point at $r_0=1$ as
shown by the dashed lines in Figure 1.
  The value of $T$ where the
curves jump separates
the two types of solutions found.
  The error bar indicates the
estimated uncertainty in the values
at $T\sim 2$ which results from
the finite grid used in calculating
$\Delta \phi(\Psi)$.
}
\label{Figure 2}
\end{figure*}

\section{Numerical Solutions of Grad-Shafranov Equation}

      We solve  equation (1)
inside a cylindrical 
``box,'' $r=0$ to $R_{m}$
and $z=0$ to $Z_{m}$ with
$H(\Psi)$ determined by iteratively
solving equations (1) - (3) 
(see Li {\it et al.} 2001;  Finn \& Chen 1990).
       The outer boundaries ($r=R_{m}$ and
$z=Z_{m}$) are treated as
conducting surfaces representing
the  interface between the coronal
${\bf B}$ field and external,
non-magnetized plasma.
   The external plasma will be pushed
outward by the coronal  field
as it is twisted.
  Thus we consider the
${\bf B}$ field behavior as $R_{m} \rightarrow \infty$
and $Z_{m}\rightarrow \infty$.

     For a dipole-like field threading
the disk  (Figure 1), it is natural
to measure lengths in terms of the radius
of the $O-$point in the disk, $r_0$.  
  The natural value of the flux function
$\Psi$ is its maximum value at the $O-$point,
$\Psi_0$. 
   In turn, a natural unit for the magnetic
field strength is $\Psi_0/r_0^2$.
 Notice that for the case of Figure 1,
$\Psi_0/r_0^2 \approx B_z(0)/10.4$.
   Further, a natural unit for the total
current (times $c/2$), $cI_{tot}/2 = (-H)_{max}$ 
is $\Psi_0/r_0$.
    Also, we measure the total 
toroidal flux and the magnetic energy,
\begin{equation}
\Phi_t = \int drdz ~B_\phi~,
\quad W_m = {1\over 4}\int rdrdz ~{\bf B}^2~,
\end{equation}
in units of $\Psi_0$ and
$\Psi_0^2/r_0$, respectively.
   For boundary conditions,
on the disk surface, $\Psi(r,0)$ is
fixed and equal to its ``initial''
value  on the disk
owing to the disk's  perfect conductivity.
   On the $z-$axis we can
take $\Psi(0,z) =0$, because
$B_r = -(1/r)(\partial \Psi/\partial z)
=0$. 
  On the conducting outer boundaries,
$\Psi(r,Z_{m})=0$ and $\Psi(R_{m},z)=0$.
Note however that in \S14 we discuss simulation
results for the case of open outer boundaries.
 A uniform $(r,z)$ grid of $200 \times 200$
was used.

  Figure 2 shows the behavior
of a set of field solutions
of  equation (1)
as the twist $T=\omega_0t$ 
is increased. 
     The nature of the solutions
changes dramatically as the twist
increases above a critical
value $T_c \approx 1.14$ rad.
   For  $T<T_c$ the nature of
the field solutions is shown in Figures
3 and 4.  
    The twisting of the field
by the differential rotation of
the disk ``pumps'' magnetic flux
and energy into the disk corona
and the field tends to ``inflate.''
  This behavior of  coronal magnetic
field loops of the Sun 
as a result of footpoint
twisting is well known from
the works of Aly (1984, 1991) and
Sturrock (1991).  
    The self-similar inflation
of a force-free field
threading a non-Keplerian disk without
outer boundaries was 
studied Lynden-Bell
and Boily (1994) and their solution
for small twists
is analogous to our low-twist solutions.
   The expansion of a force-free field
into  finite pressure external plasma
has been studied by Lynden-Bell (1996)
and Li {\it et al.} (2001) and the {\it poloidal
field} is found to fairly uniformly fill
the coronal space.

   For $T>T_c$,
a new field configuration appears
with a different topology. 
   This is shown in Figure 5 where it
is seen that a  ``plasmoid''
consisting of toroidal flux
has detached from the disk and
is separated 
 by the dashed field
line which has an $X-$point above
the $O-$point on the disk.  
   Figure 6 shows a 3D view of two
representative field lines for
the same case.
  These high-twist equilibria consist
of a region near the axis which
is {\it magnetically collimated} by
the toroidal $B_\phi$ field and
a region far from the axis, on
the outer radial boundary, which
is {\it anti-collimated} in the sense
that it is pushed against the outer
boundary.
     The field lines returning
to the disk at $r>r_0$ are
anti-collimated by the pressure
of the toroidal magnetic field.  
  The {\it poloidal field} fills only
a small part of the coronal space.
  In a purely analytical analysis,
Heyvaerts (2001) has independently
found MHD equilibria involving the
simultaneous formation of a collimated
axial jet and an uncollimated outflow.
   Figure 6 shows a three dimensional
view of sample field lines.  

   As a test of our numerical solutions
note that conservation of axial
momentum can be written as
\begin{equation}
{\bf \nabla}\cdot
({\bf T}\cdot \hat{\bf z})  =0~,
\quad {\bf T}\cdot \hat{\bf z}\equiv
{\bf B}^2{\hat{\bf z}}/8\pi -B_z{\bf B}_p/4\pi~.
\end{equation}
    Integration of equation (5) over
the ``box'' $(R_m,Z_m)$ gives
\begin{equation}
\int_0^{R_{m}}rdr~
(B_r^2+B_\phi^2-B_z^2)_{z=0}
=\int_0^{R_{m}}rdr~(B_r^2)_{z=Z_{m}}~,
\end{equation}
where the integral on the left hand
side represents the flux of momentum 
from the disk into the box and
the other integral the flux of momentum
out of the top of the box.
  This equation
is accurately satisfied by our
numerical solutions; the typical errors
are $<0.1 \%$.

\begin{figure*} 
\caption{
Poloidal field  solutions for
case of conducting boundaries for twists 
$T =0.67$ rad.  (dashed lines) and 
$T =1.1$ rad. (solid lines).  
    The initial
poloidal magnetic field is shown by the
dashed lines in Figure 1.
}
\label{Figure 3}
\end{figure*}

\begin{figure*} 
\caption{
  The top panel shows the field
line twist function ${\cal F}(\Psi)$,
with the circles indicating the derived numerical
values and the smooth curve
the theoretical dependence for a Keplerian
disk given by equation (3).  
   The case shown corresponds 
to the solid field
lines in Figure 2 where the twist is
$T=1.1$ rad.  Here, $\Psi$ is
measured in units of $\Psi_0$,
the value at the $O-$point. 
   For this case, the lower
panel shows the poloidal current
flow $-H(\Psi)$ (in units
of $\Psi_0/r_0$), with the points
indicating the numerical values
and the smooth curve an analytic
fit. 
   The initial
poloidal magnetic field is shown by the
dashed lines in Figure 1.
}
\label{Figure 4}
\end{figure*}

\section{Analytic Solution for Poynting Jet}

   Most of the twist 
$\Delta \phi$ of a field line of a Poynting
jet occurs
along the jet from $z=0$ to $Z_{m}$.
Because $-r^2 d\phi/H(\Psi) = dz/B_z$,
we have
\begin{equation}
{\Delta \phi(\Psi) \over -H(\Psi)}=
{(\omega_0 t) {\cal F}(\Psi/\Psi_0)
\over -H(\Psi)} \approx
{Z_{m} \over r^2 B_z}~,
\end{equation}
where $r^2 B_z(r,z)$ is evaluated
on the straight part of the jet
at $r=r(\Psi)$.
   In the core of the jet
$\Psi \ll \Psi_0$, we have
${\cal F} \approx 3^{9/8}
(\Psi_0/\Psi)^{3/4}$, and in
this region we can take
$
\Psi=C \Psi_0 (r / r_0)^q$, and
$ H=-{\cal K}
({\Psi_0 / r_0})
({\Psi/ \Psi_0})^s,$
where $C,~q,~{\cal K},$ and $s$ are
dimensionless constants.
   Equation (1) for the straight
part of the jet implies
$q=1/(1-s)$ and $C^{2(1-s)}=
s(1-s)^2 {\cal K}^2/(1-2s)$.
  Thus we find
$s=1/4$ so that $q=4/3$,
$C =[9/32]^{2/3}{\cal K}^{4/3}$,
and
$
{\cal K}=3^{1/8} 4({r_0 \omega_0  t
/ Z_{m}})$.

   In order to have
a Poynting
jet, we find that ${\cal K}$ must be
larger than $\approx 0.5$.
   This corresponds directly to
the condition $T> T_c$ for the
occurrence of the high-twist solutions.
  The condition arises from the fact
that there is a competition between
the build up of toroidal flux inside the box
due to twisting by the disk
which acts to increase $B_\phi$ and the
expansion of the boundaries which acts
to decrease $B_\phi$ (see \S 12).
    If the boundaries expand too rapidly
$B_\phi$ does not increase sufficiently
to give a self-collimated Poynting jet.
    For the case of uniform expansion
of the top boundary, $Z_{m}=V_z t$,
this condition is the same as
$V_z <9.2 (r_0\omega_0)$.
   For the case of Figure (5),
${\cal K} \approx 0.844$.
  The field components in the
straight part of the jet
are
\begin{equation}
B_\phi =-\sqrt{2}B_z=-\sqrt{2}
\left({3 \over 16}\right)^{1/3} {\cal K}^{4/3}
\left({\Psi_0 \over r_0^2}\right)
\left({r_0 \over r}\right)^{2/3}~.
\end{equation}
   These dependences agree approximately
with those found in numerical
simulations of Poynting jets 
(Ustyugova {\it et al.} 2000).
  On the disk,
$\Psi \approx 3^{3/2}\Psi_0(r/r_0)^2$
for $r < r_0/3^{3/4}$.
  Using this and the formula
for $\Psi(r)$
gives the relation between the
radius of a field line in the
disk, denoted $r_d$, and its
radius in the jet,
$
{r / r_0} =6.5
({r_d / r_0})^{3/2} {\cal K}^{-1}$.
Thus the power law for $\Psi$ is applicable for
$r_1 < r <  r_2$, where
$r_1 = 6.5
r_0(r_i/r_0)^{3/2}/{\cal K}$ and
$r_2=1.9r_0/{\cal K}$,
with $r_i$  the inner radius
of the disk.
  The outer edge of the Poynting
jet has a transition layer where
the axial field changes from $B_z(r_2)$
to zero while (minus) the toroidal field
increases from $-B_\phi(r_2)$
to $(-H)_{max}/r_2$.
    Using equation (6), which is
only approximate at $r_2$, gives
$(-H)_{max}\approx 1.2{\cal K} \Psi_0/r_0$, 
  This dependence agrees approximately
with our Grad-Shafranov solutions.

\begin{figure*} 
\caption{
Poloidal field lines  for Poynting jet
case  for twist 
$T =1.84$ rad. and $(-H)_{max} =1.13$.
    The initial
poloidal magnetic field is shown by the
dashed lines in Figure 1.
  The dashed contour is the separatrix
with the X-point indicated.
   Note that the radial width 
of the upgoing
field lines along the axis is about
one-half the width of downgoing field
lines at the outer wall as required
for equilibrium.
}
\label{Figure 5}
\end{figure*}

\begin{figure*} 
\caption{
Three dimensional view
of two field lines originating from
the disk at $x=\pm 0.32~r_0$
($\Psi =0.4$) for the Poynting jet
of Figure 5.
  Each field line has a twist of
$\approx 8.22$ rad. or about $1.31$
rotations about the $z-$axis from
its beginning at $r_1$ and end at $r_2$.
   The $z-$axis is tilted towards
the viewer by $30^\circ$.
}
\label{Figure 6}
\end{figure*}

\section{Expansion of Boundaries}

   The magnetic forces on the outer
wall increase by a large factor
in going from the low-twist
to high-twist solutions.
  The 
radial force on the cylindrical
wall and the axial force
on the top wall are 
\begin{equation}
F_r = {1\over 4}R_{m}
\int_0^{Z_{m}} dz~B_z^2\bigg|_{R_{m}}~,
~~ F_z = {1\over 4}
\int_0^{R_{m}}rdr~B_r^2\bigg|_{Z_{m}}~.
\end{equation}
For the low-twist solution of
Figure 3 ($T=1.1$), $(F_r,F_z) \approx
(0.0061,0.013)$,
whereas for the high-twist solution
of Figure 5 ($T=1.79$), $(F_r,F_z)\approx
(0.26,0.45)$ in units of $(\Psi_0/r_0)^2$.

   Figure 7 shows the radial dependence
of the magnetic pressure 
on the top boundary for
the cases of low-twist 
and high-twist  solutions.

 Different behavior is exhibited 
by the low-twist
and high-twist solutions as the
conducting boundaries are moved
outward.  
   For the low-twist solutions,
the poloidal field tends to
expand outward to fairly uniformly
fill the available space.
   As $R_{m} \rightarrow \infty$
and $Z_{m} \rightarrow \infty$, we
find that these solutions are
similar to those obtained 
by Lynden-Bell and
Boily (1994) where there are no
outer boundaries.

   In contrast, for the high-twist 
solutions, the poloidal field near the axis
maintains its collimation due to
$B_\phi$, whereas the ``return'' 
poloidal field far from the axis
is pushed against the outer cylindrical
wall due to the $B_\phi$ field.
     As $R_{m} \rightarrow \infty$
and $Z_{m} \rightarrow \infty$, 
the collimated field near the axis
will  extent outward along the $z-$axis
in the absence of instabilities.
    It is clear from Figure 7 that
the magnetic pressure on the top
boundary peaks near the axis. 
   Thus, this region of the boundary
should expand most rapidly in the
physical case where
the boundary is an interface
with external plasma.

    Estimation of the axial expansion
of a Poynting jet
can readily be made assuming 
a region of radius
$\sim g r_2$ ($g\sim 2 -3$) of the 
jet expands with 
velocity $V_z$ into a constant
density external medium.
   For likely conditions $V_z$
is much larger than the sound speed in
the external medium so that  
the ram pressure due to the jet motion is
$\rho_{ext} V_z^2$,
assuming $(V_z/c)^2 \ll 1$ as required
by  our non-relativistic treatment. 
  Balancing this pressure with the
magnetic pressure of the jet gives
\displaymath
\rho_{ext}V_z^2 \sim {(-H)_{max}^2 \over
4\pi g^2 r_2^2} 
\sim 0.14 [B_z(0)/g]^2 (r_0\omega_0/V_z)^4~,
\enddisplaymath
or
\begin{equation}
V_z \sim (r_0\omega_0)^{2/3}[B_z(0)/g]^{1/3}
/\rho_{ext}^{1/6}~.
\end{equation}
The condition for a Poynting jet
${\cal K}>0.5$ corresponds to 
 $V_z < 9.2(r_0\omega_0)$.
For $B_z(0)=100$G, $r_0=10^{15}$cm,
and $M=10^8M_\odot$, the external density
$\rho_{ext}$ must be larger than 
$\sim 2\times10^{-22}{\rm g/cm}^3$
in order to have $V_z <c$. 
  For larger magnetic fields, a
much larger external density
is needed to give $V_z <c$.  This
points up the need for a
relativistic treatment
of the Poynting jet expansion.

\begin{figure*} 
\caption{
   Radial dependences
of the magnetic pressure $p_{mag}=
{\bf B}^2(r,Z_{m})/4\pi$ on the top boundary for
the cases of
 high-twist ($T=1.79$) (the top curve)
and  low-twist ($T=1.1$) 
(bottom curve) solutions.
The pressure is in units of $(\Psi_0/r_0^2)^2
\approx (B_z(0)/10.4)^2$.
}
\label{Figure 7}
\end{figure*}

\section{Angular Momentum Conservation}

   Conservation of angular
momentum about the $z-$axis can
be written as
\begin{equation}
{\bf \nabla}\cdot
{\vec{\cal J}}  =0~,
\quad {\vec{\cal J}}\equiv
-r{\bf B}_p B_\phi/4\pi = - {\bf B}_p H/4\pi~,
\end{equation}
where ${\vec {\cal J}}$ is the angular
momentum flux-density vector.
    Integration of equation (11) over
the ``box'' $(R_m,Z_m)$ gives
\begin{equation}
0= -{1\over 2}\int_0^{R_{m}}rdr~H (B_z)_{z=0}~.
\end{equation}
The subscript $z=0$ here and
subsequently indicates that
the quantity is evaluated on the top
surface of the disk.
  For a dipole-like field threading
the disk where $\Psi(R_{m},0)=0=\Psi(0,0)$,
 equation (12) gives
\begin{equation}
0=-{1\over 2}\int_0^{\Psi_0} d\Psi H(\Psi)
-{1\over 2}\int_{\Psi_0}^0 d\Psi H(\Psi)~.
\end{equation}
The first integral represents the
outflow of angular momentum from the
inner part of the disk (interior to the
$O-$point), and this 
equals the angular momentum inflow into
the outer part of the disk given by the
second integral.

   The outflow of angular momentum
from the inner part of the disk
causes enhanced accretion of this
part of the disk, whereas
the inflow of angular momentum to
the outer disk reduces the accretion rate.
   Because the Poynting  outflows
carry negligible matter, the 
continuity equation for the disk
is 
\begin{equation}
\partial \Sigma/\partial t+{\bf \nabla \cdot}
(\Sigma v_r \hat{\bf r})=0~,
\end{equation}
where
$\Sigma(r,t)$ is the surface mass density
of the disk.  
  The continuity equation for the
disk angular momentum is
\begin{equation}
\partial (\Sigma {\ell})/\partial t+
{\bf \nabla \cdot}(\Sigma v_r \ell~ \hat{\bf r}
+{\cal T}_r^{vis}~\hat{\bf r})
=r(B_\phi B_z)_{z=0}/2\pi~,
\end{equation}
where ${\cal T}^{vis}_r$ represents
the viscous transport of angular
momentum in the disk, the term on the
right-hand side  
the outflow or inflow
of angular momentum (from the
two sides of the disk) due to the magnetic
field, and $\ell$ is the specific angular
momentum of the disk matter. 
  The disk is assumed
almost Keplerian so that  
$\ell =\sqrt{GM r}=rv_K$ and
consequently
the two continuity equations give
the mass accretion rate
$\dot{M}=\dot{M}_B+\dot{M}_{vis}$, where
\begin{equation}
\dot{M}_B(r) \equiv - 2\pi r\Sigma v_r=
-2(r^2/v_K)(B_\phi B_z)_{z=0} ~,
\end{equation}
is the ``magnetically driven'' mass
accretion rate, and $\dot{M}_{vis}=
6\pi\sqrt{r}~d(\nu \sqrt{r}  \Sigma)/dr$
is the viscous accretion rate
with $\nu$ the kinematic viscosity (Lovelace,
Newman, \& Romanova 1997).
  We have $\dot{M}_B > 0$ (or$ <0$)
for $r < r_0$ (or $r>r_0$).
   The accretion speed is $u \equiv -v_r=
u_B + u_{vis}$ with 
$u_B =\dot{M}_B/(2\pi r \Sigma)$ and
$u_{vis}=\dot{M}_{vis}/(2\pi r \Sigma)$.

   Figure 8 shows the radial
dependence of $\dot{M}_B$ for a
high-twist case.  
    That $\dot{M}_B$ due to
the Poynting outflow is a
function of $r$ emphasizes the fact
that disk is not  stationary.

\begin{figure*} 
\caption{
The top panel shows
the radial dependence of the mass
accretion rate of the disk $\dot{M}_B$
due to the Poynting outflow from the
disk,
and ($0.1\times$) the Poynting power outflow per
unit radius $d\dot{E}_B/dr$ for quasi-stationary
conditions.  
  The lower panel shows the radial
dependence of the angle between poloidal
field lines and the $z-$axis at the
disk surface, $\theta = \tan^{-1}(|B_r/B_z|)$.
 Both panels are for the high-twist case
of Figures 5-7.
   As mentioned, for $\Psi>0.95$ or 
$0.764 < r < 1.33$
the field lines are closed.  In the interval
$1<r< 1.5$, $B_r/B_z$ is less than zero.
}
\label{Figure 8}
\end{figure*}

\section{Magnetic Field Transport in the Disk}

    Poloidal magnetic field threading the disk
tends to be advected inward with the accretion
flow, but at the same time it may diffuse
through the disk owing to a finite magnetic
diffusivity $\eta_m$ of the disk.  
  The continuity equation for poloidal
flux through the disk $B_z(r,0,t)$ is
\begin{equation}
\partial B_z/\partial t + {\bf \nabla \cdot}
[v_r B_z \hat{\bf r} + 
{\cal U}_r B_z\hat{\bf r}]=0~,
\end{equation}
where ${\cal U}_r =(\eta_m/h) \tan(\theta)$
is the outward diffusive drift speed,
$h$ the
half-thickness of the disk,  
$\tan(\theta) 
\equiv (B_r/B_z)_{z=0}$, and a smaller, second
order diffusion term 
($\eta_m\partial^2 B_z/\partial r^2$)
has been omitted
(see Lovelace {\it et al.} 1997).
   For cases where the diffusivity
is of the order of the viscosity and
where the viscosity is given the
Shakura and Sunyaev (1973) prescription
$\nu =\alpha c_s h$ (with $\alpha <1$ and $c_s$
the midplane sound speed), the diffusive
drift speed is  ${\cal U}_r \sim \alpha c_s
\tan(\theta)$. 
   For a strong magnetic field 
threading the disk, the accretion
speed $u$ in the inner 
part of the disk ($r < r_0$)
may be large with $u_B \gg u_{vis}$ 
and $u_B > {\cal U}_r$ so that the disk
flow advects the $B_z(r,0)$ field inward.
  On the other hand at large radii,
the magnitude of $u$ is probably much smaller
so that the $B_z(r,0)$ field 
tends to drift outward, ${\cal U}_r > u$. 
 
\section{Energy Conservation}

  We assume  the coronal plasma
is perfectly conducting
so that
 ${\bf E} = -{\bf v \times B}/c$.
For
{\it quasi-stationary} and
axisymmetric conditions, ${\bf \nabla}
\times {\bf E} =0$, and thus
$
E_\phi=0,$and  ${\bf E}_p =
-{\bf \nabla}\Phi$.
so that  ${\bf v}_p \propto \pm {\bf B}_p$,
and the electrostatic
potential $\Phi= \Phi(\Psi)$. 
   Thus
${\bf E} = -\Omega {\bf \nabla}\Psi/c$
with $\Omega(\Psi) \equiv d\Phi(\Psi)/d\Psi$.
   For the situations considered here,
all field lines  pass through
the disk.  
   At the disk surface,
$E_r(r,0) = -(v_\phi)_{disk} B_z(r,0)/c$
since $v_z =0$.  
   Therefore, $\Omega[\Psi(r,0)] = 
(v_\phi)_{disk}/r$. 

 For a  force-free plasma
we have
\begin{equation}
{\partial \over \partial t}~
\left({{\bf B}^2 \over 8\pi}\right)+
{\bf \nabla}\cdot
\left( {c \over 4\pi}{\bf E} 
\times {\bf B}\right) =0~.
\end{equation}
   Integration of equation (18)
over the ``box'' ($r=0$ to $R_{m},~
z=0$ to $Z_{m}$) gives
$$
{d W_m \over dt}=
-{1\over 2}\int_0^{R_{m}} 
rdr~(v_\phi B_\phi B_z)_{z=0}
$$
$$
-{1\over 4}\dot{R}_{max}R_{m}\int_0^{Z_{m}}
dz ~{\bf B}^2(R_{m},z)
$$
\begin{equation}
-{1\over 4}\dot{Z}_{m}
\int_0^{R_{m}}rdr~{\bf B}^2(r,Z_{m})~,
\end{equation}
where $W_m$
is the magnetic field 
energy in the box (equation 4).
  The second and third integrals
in equation (19) represent the
energy expended in ``pushing'' the
boundaries outward (the $pdV$ work).  
   The first integral represents the
outflow  of energy from the inner part
of the disk surface (inside
the $O-$point) and the inflow into
the outer part of the disk.
   For a dipole-like field threading
the disk where $\Psi(R_{m},0)=0=\Psi(0,0)$,
the first integral for the power
output from the disk can be rewritten as
\begin{equation}
- {1\over 2}\int_0^{\Psi_0} d\Psi 
\Delta \dot{\phi~}(\Psi)H(\Psi)
={ \omega_0\Psi_0^2\over 2 r_0}
\int_0^1 d \tilde{\Psi} \tilde{{\cal F}}
[-\tilde{H}(\tilde{\Psi})],
\end{equation}
where $\omega_0 \equiv \sqrt{GM/r_0^3}$, and
the tildes indicate dimensionless
variables.
For the low-twist solution ($T=1.1$)
of Figure 3, the
dimensionless integral denoted
${\cal I}$  $\approx 1.75$,
whereas for the high-twist solution
of Figure 5 ($T=1.84$) 
${\cal I}\approx 3.05$.
   Thus, $\dot{E}\approx$
\begin{equation}
  0.9\times 10^{45}~
{\rm erg \over s}\left({{\cal I}\over 3}\right) 
\left({B_z(0) \over3 {\rm kG}}\right)^2
\left({r_0 \over 10^{15}{\rm cm}}\right)^{3/2}
\left( { M \over 10^8 M_\odot} \right)^{1/2}.
\end{equation}
for the power output from the two sides
of the disk. 

  The Poynting  power outflow
per unit radius from the two sides of
the disk  is 
\begin{equation}
{d \dot{E}_B / dr }= -r v_K (B_\phi B_z)_{z=0}
=\left({v_K^2 / 2 r}\right)\dot{M}_B~,
\end{equation}
for {\it quasi-stationary} conditions. 
  Figure 8  shows the radial dependence
of $d\dot{E}_B/dr$ which indicates that
most of the power outflow occurs in
the inner part of the disk.

\section{Generation of Toroidal Flux}

   Perfect conductivity 
and Faraday's law
imply 
\begin{equation}
{\partial {\bf B} \over \partial t}
= {\bf \nabla} \times ({\bf v \times B})~,
\end{equation}
where ${\bf v}$ is the plasma flow velocity.
   We apply this equation to a
``box''  extending from the disk
at $z=0$ to a non-rotating, perfectly
conducting surface at $z=Z_{m}$,
and from the axis $r=0$ to
a cylindrical, non-rotating,
perfectly conducting
surface at $r=R_{m}$.  
  These outer surfaces are allowed
to move so that $R_{m}$ and $Z_{m}$
are in general time-dependent.
  The toroidal flux $\Phi_t$ (equation 4) 
obeys
\begin{equation}
{d\Phi_{t} \over dt}\!=\!\!\int_0^{R_{m}}
\!\!dr(v_\phi B_z - v_z B_\phi)\bigg|_0^{Z_{m}}
 -\int_0^{Z_{m}}dz(v_rB_\phi -v_\phi B_r)
\bigg|_0^{R_{m}}.
\end{equation}
  On the top boundary $z=Z_{m}$,
$B_z =0$ and $B_\phi=0$; 
on the outer cylindrical boundary
$r=R_{m}$, $B_r=0$ and $B_\phi=0$;
on the $z-$axis $B_\phi=0$ and $B_r=0$;
and on the disk $v_z(r,z=0)=0$.
   Consequently, equation (24) simplifies
to 
\begin{equation}
{d \Phi_{t} \over dt}
= -\int_0^{R_{m}} dr~v_\phi B_z(r,z=0)~,
\end{equation}
where $v_\phi =\sqrt{GM/r}$ is the
azimuthal velocity of the disk. 
   The integrand of equation (25)
is  independent of time in view
of our assumption that $\Psi(r,0)$,
and  thus $B_z(r,0)=
(1/r)\partial \Psi(r,0)/\partial r$,
is time-independent. 

   For a dipole-like field threading
the disk [where $\Psi(r,0)$ increases
from zero to a maximum $\Psi_0$
and then decrease to zero at
$r=R_{m}$],
equation (25) can be rewritten as
\begin{equation}
{d\Phi_{t} \over dt}=
- \int_0^{\Psi_0} d\Psi 
~\Delta \dot{\phi~}(\Psi)~,
\end{equation}
where $\Delta \dot{\phi~}(\Psi)$ is the time
derivative of equation (3), which
is independent of time.
  For the dipole-like field of Figure 1,
evaluation of this integral gives
$d\Phi_{t}/dt 
\approx -12.1 \omega_0 \Psi_0 \approx
-12.1\omega_0 [B_z(0)/10.4]r_0^2 $.
Thus, $d \Phi_t/dt \approx$
\begin{equation}
 -1.3\times 10^{28}~
{{\rm Gcm^2}\over {\rm s}}
\left({B_z(0) \over 3{\rm kG}}\right)
\left({r_0 \over 10^{15}{\rm cm}}\right)^{1/2}
\left( { M \over 10^8 M_\odot} \right)^{1/2}
\end{equation}
is the toroidal flux generation rate from
one side of the disk.


\section{Poynting  versus Centrifugal
Outflows}

     The centrifugal
force near the surface of a Keplerian 
disk has a role
in launching hydromagnetic outflows 
(Blandford \& Payne 1982).  
   The ``centrifugal launching'' requires
that the field lines (projected into
the poloidal plane) be tilted away
from the $z-$axis by an angle $\theta$
greater than $30^\circ$.
      The magnetic force, dominantly $d(B_\phi^2/8\pi)/dz$,
is comparably important for launching hydromagnetic
outflows (Lovelace, Berk, \& Contopoulos 1991;
Ustyugova {\it et al.} 1999); it also
increases as $\theta$ increases.
    Figure 8 shows the radial variation of
$\theta(r) = \tan^{-1}(|B_r/B_z|)_{z=0}$,
for a high-twist case.
   For this case, $\theta <30^\circ$
for $r<0.3r_0$, which includes the part
of the disk giving the largest power
output per unit radius, $d \dot{E}_B/dr$,
shown in the top panel of Figure 8.
   Thus there is a Poynting outflow under
conditions where {\it no} centrifugal outflow
occurs.   

   However, for the part of
the disk where $\theta>30^\circ$
we predict a
centrifugal hydromagnetic outflow
with $-(B_\phi B_z)>0$ and inward
magnetically driven accretion, $u_B >0$
(see Ustyugova {\it et al.} 1999).
 Hydromagnetic outflows from
the region $r>r_0$ are found in the 
MHD simulations discussed by
Ustyugova {\it et al.} (2000)
and in \S 13.

\section{Collapse of Inner Disk}

    Quasi-stationary Poynting jets
from the two sides of the disk within $r_0$
give an energy outflow
per unit radius of the disk
$d\dot{E}_B/dr=
r v_K(-B_\phi B_z)_h$, where the $h$
subscript indicates evaluation at
the top surface of the disk.
   This outflow
is $\sim r_0d\dot{E}_B/dr \sim
v_K(r_0)(\Psi_0/r_0)^2$ is
estimated in equation (22), which agrees
approximately with the values
derived from the simulations (see \S 14).

      For long time-scales,
the Poynting jet is of course time-dependent
due to the angular momentum it extracts
from the inner  disk ($r<r_0$).
    This loss of angular momentum leads to
a ``global magnetic instability'' and collapse
of the inner disk (Lovelace {\it et al.} 1997).
   An approximate model of this collapse
can be made if the
inner disk mass $M_d$ is concentrated
near the $O-$point radius $r_0(t)$,
if the
field line slippage through the
disk is negligible (see \S 8),
$\Psi_0=$const, and if
$(-rB_\phi)_{max}\sim \Psi_0/r_0(t)$ (as
found here).  Then,
$
M_d {d r_0 / dt} =  {-2 \Psi_0^2
(GMr_0})^{-1/2}.$
   If  $t_i$ denotes the time at which
$r_0(t_i)=r_i$ (the inner radius of the disk), then
$
r_0(t)=r_i [1 -(t-t_i)/
t_{coll} ]^{2/3}~,
$
for $t \leq t_i$,
where $t_{coll} =\sqrt{GM}~M_d r_i^{3/2}
/(3 \Psi_0^2)$
is the time-scale for
the  collapse of the inner disk.
(Note that the time-scale for $r_0$ to decrease
by a factor of $2$ is $\sim t_i(r_0/r_i)^{3/2}\gg t_i$
for $r_0 \gg r_i$.)
  The power output to the Poynting jets is
\begin{equation}
\dot{E}(t)={2\over 3}{\Delta E_{tot}\over
 t_{coll}}\left(1-{t-t_i\over
 t_{coll}}\right)^{-5/3}~,
\end{equation}
where $\Delta E_{tot} = G M M_d /2 r_i$ is
the total energy of the outburst.
  Roughly, $t_{coll} \sim 2~{\rm day}
M_8^2(M_d/M_\odot)(6\times 10^{32}{\rm
G cm}^2/\Psi_0)^2$ for
a Schwarzschild black hole, where validity of the
analysis requires $t_{coll} \gg t_i$.
   Such outbursts may explain the flares
of active galactic nuclei blazar sources 
(Romanova \& Lovelace 1997;
Romanova 1999; Levinson 1980)
and the one-time outbursts of gamma ray burst
sources (Katz 1997).

\section{Occurrence of Poynting Jets and
Kink Instability}

   The rate at which toroidal flux $\Phi_t$
is created in the region above
the disk is $d\Phi_t/dt \approx -12\omega_0 \Phi_0$
(see \S 10).  
     If the area $R_mZ_m$ of this region is fixed
owing to a  very dense external plasma, 
the average value of $-B_\phi>0$ in it
 increases.
    On the other hand if the area 
$R_mZ_m$ increases rapidly enough,
$-B_\phi$ will decrease.  
    The Poynting jets occur under  conditions where
$-B_\phi$ increases to a sufficient extent
to cause pinching of the poloidal magnetic
field.
     Because the most rapid expansion of
the ${\bf B}$ field of occurs
in the $z-$direction, a necessary condition for
occurrence of a  Poynting jet is that
the rate of expansion of the boundary
$V_z=d Z_m/dt$ be bounded by some constant.
   In fact  the  condition obtained in \S 4 for 
occurrence of Poynting jets  has this form, 
$V_z < 9.2(\omega_0 r_0)$.

   Note that there may be ``self-regulation''
in the respect that the field configuration
which occurs is at the ``boundary'' between 
low and high-twist solutions.  
   In view of Figure 7, the low-twist field
gives a gradual expansion of the boundaries
which allows build up of toroidal flux
whereas the high-twist field gives a more rapid
expansion which tends to give a slower
increase of $\Phi_t$.

   The region of collimated field (see Figure 5) -
the Poynting jet - has ${\bf v}^2 \ll v_A^2$
and is kink unstable according to the standard
non-relativistic analysis (e.g., Bateman 1980,
ch. 6).
   The instability will lead to a helical 
distortion of the jet with the non-linear amplitude
of shift  of the helix $\Delta r \sim v_A t$ and
with the helix having the same twist about the
$z-$axis as the axisymmetric ${\bf B}$ field.
  Note however that for the astrophysical
conditions of interest $v_A \equiv |{\bf B}|
/\sqrt{4\pi \rho}$ is likely to be larger than
the speed of light.  A relativistic perturbation
analysis is then required including the displacement
current. The physical Alfv\'en speed is
$V_A=c/(1+c^2/v_A^2)^{1/2} < c$.
   Thus the speed  of lateral displacement of
the  helix is less than c.
    The evolution of the Poynting
jet evidently depends on {\it both}
$V_A$ and the velocity of propagation
of the ``head'' of the jet $V_z$ (\S 5)
which may be relativistic.
   Relativistic propagation of the jet's
head may act to limit the amplitude
of helical kink distortion of the jet.
   On the other hand a sub-relativistic
propagation of the head may allow the
helix amplitude to grow but this amplitude
can be  limited by flux  conservation
as discussed by Kadomtsev (1963).

\section{MHD Simulations of Poynting Jets}

   For the MHD simulations described here,  
the initial  magnetic field has
a dipole-like form as shown in
Figure 1.
    The computational region  
$r=0$ to $R_{max}$, $z=0$ to $Z_{max}$
is taken to have $R_{max}=Z_{max} \approx 10r_0$.
  Initially, the corona of the disk is in 
isothermal equilibrium without
rotation.
   At $t=0$ the disk starts to 
rotate  with Keplerian velocity  
$v_\phi(r,0)= r\Omega_K $, where
$\Omega_K = \sqrt{GM}/(r^2+r_i^2)^{3/4}$,
where the smoothing length $r_i =0.2r_0$ is
interpreted as the inner radius of the disk.
  The smoothed gravitational potential
is $-GM/(r_i^2+{\bf r}^2)^{1/2}$.

\begin{figure*} 
\caption{
Time evolution of dipole-like
field threading the disk from the
initial configuration $t=0$ (bottom panels)
for the case of conducting outer boundaries
at $R_m=10$ and $Z_m=10$.  
   Here, $t_i$ is the period of rotation
at the inner radius of the disk $r_i$.
  The initial field is shown by the dashed
lines in Figure 1.
   The left-hand
panels show the poloidal field lines which
are the same as $\Psi(r,z)=$const lines.
  The right-had panels show the poloidal
velocity vectors ${\bf v}_p$.  
   For this calculation a uniform $100 \times
100$  grid was used.  For times longer
than $\sim 8t_i$ the numerical calculations
crash apparently because of the build up
of fine scale structure in the simulation
region.
}
\label{Figure 9}
\end{figure*}

  On the disk surface, the boundary
conditions  are as follows
(Ustyugova {\it et al.} 2000).
   Two of the boundary conditions 
come from the fact that
the tangential electric field $({\bf E}^\prime)_t$ 
in the frame
rotating with the disk (at the
Keplerian velocity) is zero;
$B_z$ at the disk surface is time-independent
whereas $B_r$ and $B_\phi$ at
the surface vary with time.
  Two further boundary conditions fix the
entropy of the plasma coming out of
the disk to be $s_d(r)$ and   
the density of the outflowing
plasma to be $\rho_d(r)$.
   If $v_z$ at the disk surface,
calculated by solving the MHD
equations in the computational region, 
increases to the point where it is
larger than the slow magnetosonic speed in
the $z-$direction at the disk's
surface  $c_{smz}$, then
we clamp it to be equal to $c_{smz}$.
     This
condition represents a
limit on the mass efflux
$\rho v_z$ from the disk. 
   For sub-slow magnetosonic
outflow from the disk $v_z < c_{smz}$, we have 
four boundary conditions, whereas when
$v_z=c_{smz}$ we have five boundary conditions.

  For the outer 
boundaries, we first consider the
case where these surfaces are
perfect conductors.
  Secondly, we consider the case of
 ``free'' outer boundaries where
$\partial F_j/\partial n =0$ on all scalar
variables {\it except} for the toroidal magnetic
field.  
  For this field component we take
$[{\bf B}_p \cdot 
{\bf \nabla}(rB_\phi)] =0$ on the
outer boundaries, which
was shown by Ustyugova {\it et al.} (1999) to
avoid artificial collimation which can
come from using the ``free'' boundary condition
on $rB_\phi$.
   The free outer boundary conditions
allow matter and Poynting
flux to freely flow out through these surfaces.

   For the cases we discuss, 
the strength
of the poloidal magnetic field at
the inner radius of the disk
corresponds to $(v_{Ap}/v_K)_i=16.5$
and $(c_s/v_K)_i=1$, where
$v_{Ap} \equiv |{\bf B}_p|/\sqrt{4\pi \rho}$.
  The $i-$subscript indicates evaluation
at the inner radius
of the disk $r=r_i$ on the disk surface.
   In the midplane of the disk $(v_{Ap}/v_K)_{z=0}$
is less than or much less than unity.
    Different radial profiles of $c_s$ on
the disk surface have been used with
similar results,
including $c_s/v_K=$ const 
and $c_s=$ const;  the density profiles
on the disk surface have  been obtained
as in Ustyugova {\it et al.} (1999).

    Figure 9 shows the evolution of
the coronal plasma for the case of fixed,
conducting outer boundaries at $(R_m,Z_m)$.
   After about $6$ rotation periods of 
the inner disk, the outgoing  poloidal field 
collimates along the $z-$axis, and the returning
poloidal field is pressed outwards to the
conducting walls.  Most of the magnetic
field  is strongly
field-dominated with flow speeds ${\bf v}^2
\ll { v}_A^2$, where $v_A$ is the Alfv\'en
velocity.
    The field configuration is similar to that
found for the high-twist Grad-Shafranov solutions
as shown in Figure 5.

   Figure 10 shows the long time evolution of
the coronal plasma for the case where the outer
boundaries are  free  (i.e.,  open)
boundaries.  
   These simulations evolve to
a quasi-stationary final state where most
of the region is strongly field-dominated.
    In the jet region along the $z-$axis,  
the poloidal field is collimated by the
$B_\phi$ field.
   We find that the 
profiles of $B_\phi(r)$ and $B_z(r)$ from
the simulations (Ustyugova {\it et al.} 2000)
agree to a  good approximation with 
equation (6) of our analytic model.
  The main respect in which the simulations
differ from the Grad-Shafranov solutions
is that region close to the disk where
the magnetic field returns to the
disk is {\it not} field-dominated.
  Instead this region has a hydromagnetic
outflow from disk.
    This is predicted
theoretically (Blandford \& Payne 1982)
and observed in simulations (Ustyugova {\it et al.} 1999)
to be the case when the
angle $\theta$ between the disk normal and the
poloidal field lines is larger than $30^\circ$.

\begin{figure*} 
\caption{
   Time evolution of dipole-like
field threading the disk from the
initial configuration $t=0$ (bottom panels)
to the final quasi-stationary for
the case of {\it open} outer boundaries
at $r=R_m$ and $z=Z_m$.
Here, $t_{out}$ is the rotation period of
the disk at the outer radius $R_{m}$ of
the simulation region; 
   for the parameters used, $t_{out}
\sim 200 t_i$.
  The initial field is shown by the dashed
lines in Figure 1.
    The left-hand
panels show the poloidal field lines which
are the same as $\Psi(r,z)=$const lines;
$\Psi$ is
normalized by $\Psi_0$ (the maximum
value of $\Psi(r,z)$) and the
spacing between lines is $0.1$.
  The middle panels show the poloidal
velocity vectors ${\bf v}_p$.
   The right-hand panels show the constant
lines of $-rB_\phi(r,z)>0$ in
units of $\Psi_0/r_0$ and the spacing
between lines is $0.1$.
  For this calculation a  $100\times 100$  
inhomogeneous grid was used with  $\Delta r_j$
and  $\Delta z_k$  growing with distance $r$   
and  $z$ geometrically as $\Delta r_j=\Delta r_1 q^j$
and $\Delta z_k = \Delta z_1 q^k$,
with $q=1.03$ and $\Delta r_1=\Delta z_1 =0.05r_0$
(Ustyugova {\it et al.} 2000).
}
\label{Figure 10}
\end{figure*}

\section{Conclusions}

    An ordered magnetic field threading
an accretion disk can give
powerful outflows or jets of matter,
energy, and angular momentum.
   Most of the studies  have
been in the hydromagnetic
regime where there is
appreciable mass outflow and find
asymptotic flow speeds of the
order of the maximum Keplerian
velocity of the disk, $v_{Ki}$.
   These flows are clearly relevant
to the jets from protostellar systems
which have flow speeds  of the order of $v_{Ki}$.
    In contrast,
observed VLBI jets in quasars and
active galaxies point to bulk
Lorentz factors $\Gamma \sim 10$ - much
larger than the disk Lorentz factor.
   In the jets of gamma ray burst 
sources, $\Gamma \sim 100$.
   The large Lorentz factors
as well as the small Faraday rotation
measures suggest that these
jets are in the Poynting flux regime.
    This work presents self-consistent
solutions for the axisymmetric,
non-relativistic plasma equilibria described
by the force-free Grad-Shafranov equation.
    We find solutions with {\it magnetically
collimated} Poynting jets 
where there is a {\it continuous} outflow
of energy, angular momentum, and toroidal magnetic
flux from the disk into the external space.  
   This behavior contradicts
the commonly accepted ``theorem'' 
of Solar plasma physics that the motion
of the footpoints of a magnetic loop structure leads
to a stationary magnetic field configuration 
with zero power, angular momentum,  and flux outflows
(Aly 1984, 1991).

   Important issues remain to be investigated -
the relativistic expansion of the  head of
a Poynting jets into an external medium and
the three-dimensional kink instability of
the jet.  Further, the magnetic extraction
of energy from a rotating black hole may be  important
(Blandford \& Znajek 1977;  Livio, Ogilvie, 
\& Pringle 1999).

  We thank 
S.A.  Colgate and J.M. Finn for valuable discussions.
   One of us (RL) thanks J. Chen for
for valuable discussions about Coronal
magnetic fields.
     This research was
supported in part by 
NASA grants NAG5-9047 and NAG5-9735 
and NSF grant AST-9986936.
MMR received partial support
from NSF POWRE grant
AST-9973366.

\end{document}